\documentclass[12pt]{spieman}  
\usepackage{amsmath,amsfonts,amssymb}
\usepackage{graphicx}
\usepackage{setspace}
\usepackage{tocloft}
\usepackage{lineno}
\title{Investigating the role of magnetic fields in the formation
and evolution of striations in interstellar clouds with PRIMA}

\author[a, e, *]{Raphael Skalidis}
\author[b, c]{Konstantinos Tassis}
\author[d]{Aris Tritsis}
\author[e]{Paul F. Goldsmith}
\affil[a]{TAPIR, California Institute of Technology, MC 350-17, Pasadena, CA 91125, USA}
\affil[b]{Department of Physics, University of Crete, Voutes Campus, GR-70013, Heraklion, Greece}
\affil[c]{Institute of Astrophysics, Foundation for Research and Technology-Hellas, Vasilika Vouton, GR-70013 Heraklion, Greece}
\affil[d]{Institute of Physics, Laboratory of Astrophysics, Ecole Polytechnique Fédérale de Lausanne (EPFL), Observatoire de Sauverny,
1290, Versoix, Switzerland}
\affil[e]{Jet Propulsion Laboratory, California Institute of Technology, 4800 Oak Grove Drive, Pasadena, CA 91109-8099, USA}

\cftpagenumbersoff{figure}
\cftpagenumbersoff{table} 
\begin{document} 
\maketitle

\begin{abstract}
Striations are diffuse, linear, quasi-periodic, and magnetized structures located in the outskirts of molecular clouds. These structures seem to play an important role during the earliest stages of star formation. Theoretical models suggest that magnetic fields play an important role in the formation of striations. With its unprecedented resolution and sensitivity, the polarization module of the PRIMAger instrument onboard the PRIMA space observatory will enable studies related to the magnetic properties of striations in nearby molecular clouds. We plan to target three nearby ($\lesssim 350$ pc) molecular clouds (the Polaris Flare, Taurus, and Musca) with prominent striations that are strongly coupled to the large-scale magnetic field properties, as traced by low-resolution sub-millimeter polarization data. We will search for the unique imprint of the passage of magnetohydrodynamic waves in the polarization angle maps, which traces the magnetic field morphology, in bands 3 (172$\mu$m) and 4 (235$\mu$m) of PRIMAger. Each of the target regions is approximately 1 square degree in size. All three regions combined, can be mapped to more than five-sigma detection in averaged polarized intensity in $\sim 59$ hours. The proposed survey promises to provide important information on the early phases of star formation.
\end{abstract}


{\noindent \footnotesize\textbf{*} Hubble Fellow, \linkable{skalidis@caltech.edu} }

\begin{spacing}{2}   

\section{Introduction}
\label{sect:intro} 

\subsection{The multiphase interstellar medium}

The interstellar medium (ISM) is a complex and dynamic environment characterized by a multiphase structure. The ISM structures are composed of gas in different phases, each with its own distinct physical properties (e.g., temperature, density). The various ISM phases dynamically interact, usually through mass or energy exchange, with the diffuse phases usually operating as a mass reservoir for the dense, cold phases, where star formation occurs\cite{valdivia_2016.H2.formation.simulations}. 

The last decade has been transformational for studies of the magnetic field in the ISM. Observations revealed that structures of the cold, neutral, and hot, ionized ISM strongly correlate with the ISM magnetic field \cite{clarck_2014.hi.fibers, kalberla_2016.hi.fibers,Bracco20, skalidis_2022.ursa.major}. The magnetic field is a dominant player in this new picture of the ISM that has emerged from observations by the Planck satellite \cite{planck_xxxv_2016}, influencing all major components and processes taking place in the ISM, including the transition of hot, diffuse gas to dense, cold material \cite{skalidis_2022.ursa.major}.

Polarization studies have focused on the cold and dense phase of the ISM, primarily due to the increase of infrared and sub-mm data, from observatories with unprecedented angular resolution (e.g., SOFIA, JCMT, and ALMA). As a result, polarization studies of the ISM have shifted focus, from mapping the diffuse medium to resolving the (smallest) densest structures, where protostars form. However, due to their limited sensitivity, these observatories require a considerable observational time to detect the weaker signal in diffuse ISM regions. Polarized emission data from the Planck satellite have advanced magnetic field studies of the diffuse ISM, but at a very low resolution, with an averaging of at least $20'$ required to obtain significant detections \cite{skalidis_2019}. These limitations have impacted studies of the magnetic field in the diffuse parts of ISM molecular clouds, which require observations with high dynamical range. \textit{PRIMA\cite{glenn_2024.prima.concept} offers a unique opportunity to fill the knowledge gap between the diffuse, magnetized ISM and its relation to dense, cold structures with star formation activity.}  

\subsection{The interaction between filaments and striations}

The Herschel observatory has made significant contributions to our understanding of star formation. Data from Herschel showed that nearby molecular clouds, independently of their densities and star formation rates, are comprised of dense and cold filamentary structures \cite{Andre10}. To distinguish from the general use of the term, we use {\it filaments} to refer to elongated molecular structures observed in nearby ($\lesssim 400$ pc) molecular clouds with an aspect ratio of at least 3:1 \cite{panopoulou_2017.filament.width} and column densities of the order of $10^{21} - 10^{22}$ cm$^{-2}$ \cite{hacar_2023.filament.review}. Starless and protostellar cores are usually embedded within these filaments \cite{konyves_2015}.  

Filaments are magnetized and can form by the stochastic motions of compressing gas \cite{federrath_2021, inoue_2018.filament.formation} or due to ion-neutral friction \cite{Hennebelle13}. Several key questions are still open regarding the formation and interaction of filaments with ambient diffuse structures in molecular clouds. The diffuse parts of molecular clouds, in turn, often exhibit a morphology characterized by structures known as striations \cite{Goldsmith08}.

Striations are elongated hair-like structures, observed in the low--column--density outskirts of molecular clouds. These structures bare similar morphological and magnetic field properties to cirrus clouds \cite{tritsis_2019}, as seen in the images of the Infrared Astronomical Satellite \cite{Boulanger88}. Striations were first observed in the J=1-0 emission of $^{12}$CO and $^{13}$CO in the northwest part of the Taurus molecular cloud, where they do not appear to be intimately associated with any denser structures \cite{Goldsmith08}. These features extend over a velocity range of 2 km s$^{-1}$, but exhibit sub-structures over narrower velocity intervals. The width of the striations was marginally resolved by the CO spectroscopic data. 

The surface brightness increments of striations are 15 - 25 $\%$ above the background CO total emission over the full velocity interval but larger increments are also observed (50-100 $\%$) over small velocity ranges with kinematically individual features. Striations were also observed by Herschel in dust emission. One of the most representative examples is the Polaris flare where well-ordered, low-density elongations are seen throughout the cloud\cite{Miville-Deschenes10, Hennemann12, Palmeirim13, AlvesdeOliveira14}. Striations have been found by Herschel in several nearby molecular clouds \cite{AlvesdeOliveira14} and they have properties similar to those of diffuse, atomic clouds having linear morphologies \cite{clarck_2014.hi.fibers, kalberla_2016.hi.fibers,tritsis_2019,skalidis_2022.ursa.major}. Striations are found to be aligned with the plane-of-sky component of the local magnetic field revealed by polarization observations. The alignment between these structures and the magnetic field has been pointed out in all studied clouds \cite{Goldsmith08, Chapman11, Hennemann12, Palmeirim13, AlvesdeOliveira14, panopoulou_2016.PFstriations}.

There is little doubt that the formation mechanism of striations is linked to magnetic fields \cite{Goldsmith08, Heyer16}. Striations were modeled as density fluctuations associated with magnetosonic waves in the linear regime \cite{Tritsis16}. These waves are excited as a result of the passage of Alfvén waves, which couple to other magnetohydrodynamic (MHD) modes through phase mixing. In contrast, it has been proposed that striations do not represent real density fluctuations but are rather a line-of-sight column-density effect due to oblique shock interactions \cite{Chen17}. However, recent evidence of enhanced $^{12}$CO formation in striations indicates that these structures represent real density enhancements \cite{skalidis_2023.striations}. Striations have also been interpreted in the context of the Kelvin-Helmholtz instability (KHI) \cite{Heyer16}, and as sub-Alfvénic streamlines in which material flows into or out of denser filaments \cite{Li13}. However, numerical experiments suggest that these mechanisms are unlikely since they do not explain the observed contrast in the column density maps of striations \cite{Tritsis16}.

High angular resolution and sensitivity polarimetric imaging data would be of great interest to set direct observational constraints and test theoretical predictions. Specifically, the magnetosonic wave model predicts that a zoo of MHD wave effects should be observable in these regions.Such waves include the ``sausage'' mode (Figure~\ref{fig:sausage_cartoon}), which has been extensively studied in the context of heliophysics \cite{Nakariakov16} and could open a new window to probe the local conditions in ISM clouds \cite{Tritsis18}. 

The polarization module of the PRIMAger \cite{ciesla_2024.primager, burgarella_2024.primager} instrument onboard the PRIMA space observatory is ideal for this endeavor. The target angular resolution of PRIMAger will allow us to study, for the first time, the small-scale fluctuations in the density and the magnetic field of striations. Such a study has not been feasible until now because past polarization observatories, such as Planck, could not probe the spatial scales of striations (\S~\ref{sec:theory}). Starlight polarization can achieve the necessary resolution but limitations arise due to the sparse sampling of the data. 

With PRIMAger we will be able to explore the correlation of the column density and polarization maps of striations to search for the unique imprint of the passage of the various wave modes. We propose to target three ISM clouds with prominent striations: the Polaris Flare, Taurus, and Musca. The striations of these clouds have been studied extensively in the past and there is already a wealth of archival data, such as emission lines of $^{12}$CO (including isotopologues), and stellar polarization data, that will allow us to build a comprehensive picture of striations and how they shape the early stages of star formation.

\section{Formation mechanism of striations}
\label{sec:theory}

\begin{figure}
\begin{center}
\begin{tabular}{c}
\includegraphics[height=8.5cm]{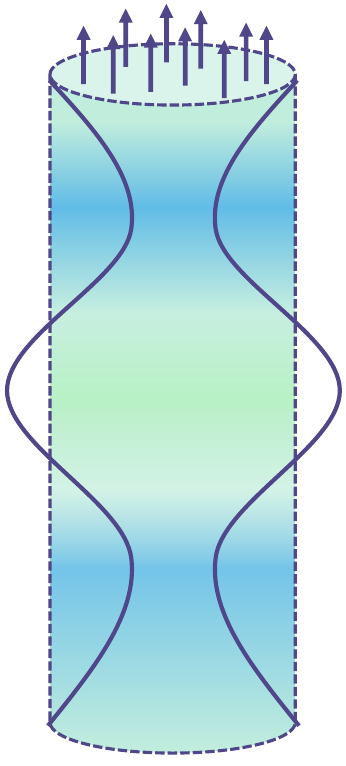}
\end{tabular}
\end{center}
\caption 
{\label{fig:sausage_cartoon} Schematic representation of sausage mode oscillations. A cylindrical flux tube of magnetic field lines is perturbed, exciting sausage waves that propagate along the mean field. The pattern of sausage modes is characterized by narrowing and widening of the magnetic field lines, as depicted by the solid black curves. Tmonashe color differences indicate changes in magnetic field strength, where blue and green areas denote higher and lower values of magnetic field intensity, respectively.}
\end{figure} 

Linear MHD waves have been proposed as a formation mechanism of striations\cite{Tritsis16}. A main prediction of this theory -- the existence of normal modes in isolated clouds -- has been observationally confirmed \cite{Tritsis18}. This was a major step for the linear MHD theory, underscoring its predictability power in observations. Another prediction of this theory, which we plan to test with PRIMager's polarization data, is the existence of coherent-structure compressible fluctuations.

Energy in MHD turbulence is distributed between waves and coherent structures \cite{yang_2019.energy.occupation}. Waves are perturbative solutions to the MHD equations and satisfy dispersion relations, such as those of the slow and fast magnetosonic modes, and the Alfv\'enic mode. Coherent structures are localized fluctuations with a high degree of spatial regularity and can be found across a range of scales. Strongly magnetized MHD turbulence is characterized by axially symmetric fluctuations \cite{goldreich_sridhar_1995}, which favor the formation of cylindrical coherent structures; sheet-like structures arise near the dissipation scales due to the alignment of magnetic and velocity fluctuations\cite{boldyrev_2006.dynamic.alignment}. Sausage modes are also linear solutions to the MHD equations that describe axisymmetric perturbations of cylindrical magnetic flux tubes, and satisfy a dispersion relation \cite{roberts_1984.sausage.modes}. Fig.~\ref{fig:sausage_cartoon} shows the oscillatory pattern of sausage modes, which is characterized by compressions and rarefractions of the magnetic field lines. Sausage-modes have been observed in the solar corona \cite{nakariakov_2003.sausage.first.detection}, where magnetic pressure is dominant compared to the thermal pressure of the gas. Similar conditions apply to striations \cite{skalidis_2023.striations, panopoulou_2016.PFstriations}, thereby maximizing the probability of detecting such modes.

\begin{figure*}
\begin{center}
\begin{tabular}{c}
\includegraphics[width=0.95\textwidth]{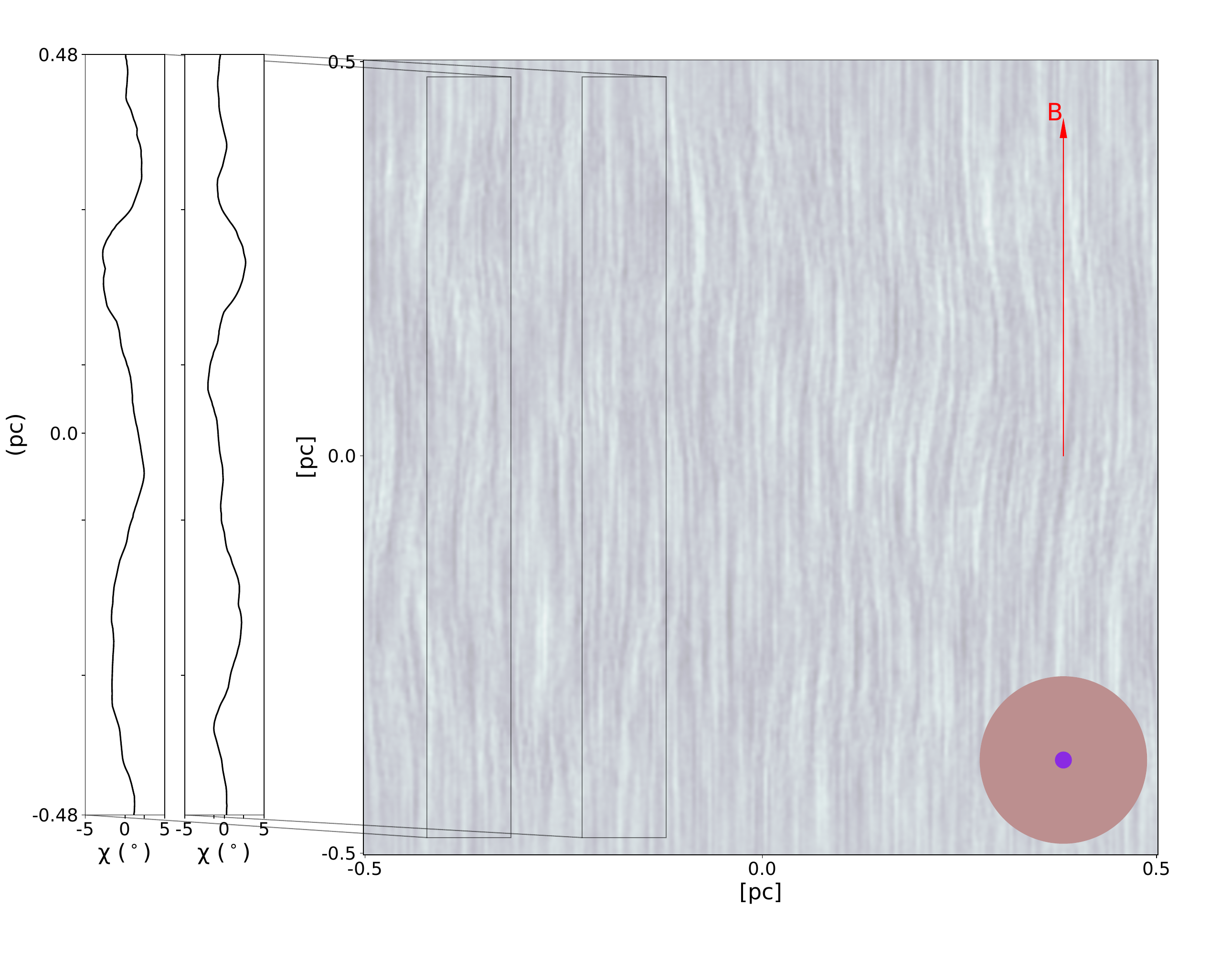}
\end{tabular}
\end{center}
\caption 
{\label{fig:striations} Striations in a 3D MHD simulation of a supersonic ($\mathcal{M}_{s}$=2) and sub-Alfvenic ($\mathcal{M}_A$=0.75) cloud \cite{Tritsis16}. For visualization clarity, we only show the magnetic field morphology (drapery pattern\cite{King18}), as would be traced by PRIMAger at 235 $\mu$m. The beam of PRIMAger is shown as a magenta circle in the bottom right of the figure, while the native resolution of Planck, which is insufficient for the sausage mode detection, is shown as a red circle. Inset panels correspond to averaged polarization angle profiles across two different orientations along the mean field, showing what the imprint of a sausage mode propagation in the polarization angle maps might look like. PRIMAger's capabilities uniquely fit for the discovery of these patterns in ISM striations.}
\end{figure*}

Our numerical models of compressible MHD turbulence show prominent sausage-mode oscillations. Fig.~\ref{fig:striations} shows the plane-of-the-sky magnetic field morphology of a direct numerical MHD simulation with a setup similar to Ref.~\citenum{Tritsis16}. The sausage mode propagation is imprinted in the morphology of the magnetic field as traced by polarization angle measurements. The inset panels in Fig.~\ref{fig:striations} show two averaged polarization angle profiles convolved to PRIMAger's resolution at 235 $\mu$m, which is $27.6''$. The fluctuations between the two polarization angle profiles are in phase, as expected from sausage-mode oscillations (see Fig.~\ref{fig:sausage_cartoon} for comparison).

Using our reference simulation, whose properties are representative of ISM striations (e.g., sonic and and Alfv\'en Mach numbers of $\mathcal{M}_{s}$ = 2 and $\mathcal{M}_A$ = 0.75), we derive that the amplitude in the polarization angle map induced by sausage modes is $\sim 4^o$. To statistically detect such low-amplitude variations, we require a polarization angle uncertainty of the order of $\sim 1^o$. The wavelength of the sausage mode in our synthetic data is $\sim 0.5$ pc, which would be traceable by any PRIMA band in molecular clouds.

There are several challenges that have hindered the detection of sausage modes in the ISM as past and current observatories lack the sensitivity and resolution to probe the predicted fluctuations. The detection of sausage mode oscillations in the ISM can be only achieved with an instrument like PRIMAger, which can provide continuum polarization angle maps of nearby molecular clouds with high angular resolution and a signal-to-noise ratio (SNR) close to 20 -- required to reach the target uncertainty in polarization angles \cite{king_2014.robopol}.

\begin{figure}
\begin{center}
\begin{tabular}{c}
\includegraphics[height=8.5cm]{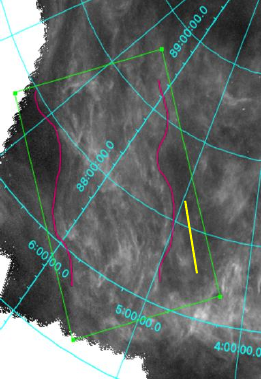}
\end{tabular}
\end{center}
\caption 
{\label{fig:polaris_flare} Striations in the Polaris Flare. The green rectangles corresponds to the proposed target region. The background maps shows the dust emission structures at 350 $\mu$m, as observed with the SPIRE instrument onboard the Herschel observatory. The yellow segment shows the average orientation of the striated structures, which are aligned with the plane-of-the-sky magnetic field morphology, as traced by starlight polarization \cite{panopoulou_2016.PFstriations}. Magenta curves mark the outline of striations, which show an oscillatory pattern similar to sausage modes.} 
\end{figure}

With the obtained polarization from PRIMAger, we would explore the correlations between total intensity (Stokes I) and polarization maps (Stokes Q and U). Under the flux-freezing approximation, magnetic field lines move in phase with the gas (ideal-MHD approximation). As a result, the characteristic features of sausage modes are imprinted in Stokes I maps as well. Dust intensity observations of the Polaris Flare at $350 \mu$m are consistent with this picture, showing a characteristic widening and narrowing of the dust structures along the mean field direction (Fig.~\ref{fig:polaris_flare}). PRIMAger's data will allow us to explore if the magnetic field morphology is consistent with the observed dust structure variations, which would be a direct evidence of a sausage mode propagation.
 
The detection of small-scale ($\sim 0.1$ pc) modes would be more challenging due to limitations arising from projection effects, instrumental noise, and interactions with other waves. To overcome these difficulties, we will employ dedicated algorithms to identify the spatial fluctuations of linear structures in striation regions \cite{sousbie_2011.disperse.algorithm, panopoulou_2014.taurus.filaments}. Power spectra of the polarization angles and linear structure intensities along the magnetic field will allow us to search for Fourier modes that satisfy the sausage mode dispersion relation at different scales. We will also search for the imprints of other linear MHD modes that could be present in the data by producing synthetic data and performing a comprehensive comparison to observations.

\section{Description of observations with PRIMAger}
\label{sec:observations}

We propose to image the polarized thermal dust emission of three $\sim 1$ square degrees areas towards the Polaris Flare, Musca, and Taurus molecular clouds using PRIMAger in order to investigate the polarization properties of striations and better understand the nature of MHD turbulence in molecular clouds. The target regions have been selected as they all exhibit prominent striations, and collectively span a wide range of ISM environments. The Polaris Flare is a diffuse molecular cloud with no star formation \cite{WardThomson_2010.polaris.cores}; Musca is on the verge of initiating star formation \cite{myers_2017.musca}, while Taurus is an active star-forming cloud \cite{hartmann_2002.taurus.sfr}. Additionally, there is a wealth of archival spectroscopic data from radio observatories that would foster valuable synergies with PRIMA \cite{Goldsmith08,panopoulou_2016.PFstriations,bonne_2020.musca.co.obs,skalidis_2023.striations}.

\subsection{Striations in the Polaris Flare}
\label{sec:polaris_flare}

In the striations of the Polaris Flare turbulence is sub-Alfv\'enic and compressible wave modes, propagating along the magnetic field lines, tend to enhance the formation of CO gas \cite{skalidis_2023.striations}. Fig.~\ref{fig:polaris_flare} shows the striations region in the Polaris Flare, where the mean orientation of the magnetically-aligned striations\cite{panopoulou_2016.PFstriations}, is shown as a yellow line. The green rectangle shows the region we propose to survey with PRIMAger. Magenta curves outline the shape of the striation region, in line with the oscillatory pattern expected from sausage modes (refer to Figs.~\ref{fig:sausage_cartoon} and \ref{fig:striations} for comparison). 

The narrowing and widening of the striations' outline (magenta curves in Fig.~\ref{fig:striations}) is $\sim 3.8$ pc, and $\sim 4.3$ pc respectively, assuming that the cloud is located at $\sim 350$ pc \cite{panopoulou_2022.cloud.distances }. The linear MHD theory predicts that the variance in the cross section of sausage modes is proportional to the fluctuating-to-ordered magnetic field ratio \cite{grant_2015}, which is, to first order, proportional to the dispersion of polarization angles. The variance in the cross section of the Polaris Flare striation's outline is $\sim 0.075$. We predict that the variance in the polarization angle along the depicted outline is $\sim 4^o$, which is consistent with our numerical simulations (Sect.~\ref{sec:theory}). Thus, our target sensitivity in polarization is SNR $\sim 20$. PRIMAger data would allow us to test this prediction by constructing the polarization angle profiles along the striations' main orientation.

Sausage modes can be only excited above a wavelength cutoff, which depends on the magnetic variance of the region \cite{edwin_roberts1983.wave.propagation.magnetic.cylinders, nakariakov_2003.sausage.first.detection}. If the presence of a sausage mode is confirmed with PRIMAger data, it would place constraints on the magnetic field strength of the region. Combined with two additional constraints, which can be obtained with spectroscopic \cite{tritsis_2018.new.method} and polarization data \cite{Skalidis2021High-accuracyPolarization, skalidis_2021.sqrt}, would enable the first on-sky cross-calibration of the various magnetic field strength estimation methods, which would provide an accurate template for ISM magnetic field studies. Finally, from Bernoulli's principle we also expect that the velocity profile will be in anti-phase with respect to the column density profiles. We plan to test the aforementioned predictions, allowing us to build a comprehensive picture for the propagation of MHD waves in the ISM.

\subsection{Striations in Musca}
\label{sec:musca}

\begin{figure}
\begin{center}
\begin{tabular}{c}
\includegraphics[height=7.5cm]{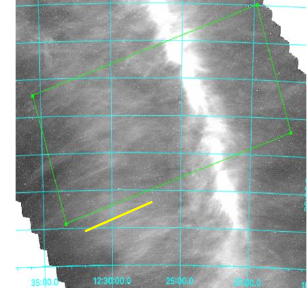}
\end{tabular}
\end{center}
\caption 
{\label{fig:musca} Striations in Musca. Striations are perpendicular to the main filament region, likely channeling material towards it. Colored shapes have the same meaning as in Fig.~\ref{fig:polaris_flare}.} 
\end{figure} 

Fig.~\ref{fig:musca} shows the dust intensity structures at 350 $\mu$m of the proposed region in Musca from Herschel. There striations are perpendicular to the main cloud structure, likely channeling material onto the cloud \cite{cox_2016}. Striations in Musca pulsate at the characteristic frequencies of normal modes, underscoring the presence of trapped MHD waves \cite{Tritsis18}. Numerical simulations reproduce the observed normal mode frequencies only for a sheet-like geometry \cite{Tritsis18, tritsis_2022.musca} and not filamentary \cite{cox_2016}, as the cloud is projected in the plane of the sky. The analytical normal mode relation suggests that the depth of Musca is $\sim 6$ pc \cite{Tritsis18}. 

The relative orientation between the projected magnetic field morphology and column density of the cloud depends on the 3D shape of the cloud. If Musca were a filament, then we would expect that the magnetic field morphology would transition from being parallel to perpendicular to the column density, as we move from the striations region to the dense part of the cloud. We refer to Fig.~4 of Ref~\citenum{andre_2019} for a visualization of this model. On the other hand, the sheet-like model, which is supported by the MHD normal mode analysis, predicts that the magnetic field geometry would smoothly transition from the diffuse (striations) to the dense part \cite{Tritsis18}. The polarization angle difference between the two models is striking, and thus even a modest accuracy in polarization angles ($\sim 5^o$) would suffice to distinguish between the two models.

PRIMAger's high angular resolution will allow us to probe potential changes in the magnetic field morphology across the main structure of Musca, which is $\sim 0.1$ pc -- Musca is located at $\sim 170$ pc\cite{tritsis_2022.musca}. The pixel size of PRIMAger's longest-wavelength band is $27.6''$, corresponding to $\sim 0.023$ pc, which is sufficient to trace any transitions in the morphology of the magnetic field across Musca. The proposed polarization observations would provide important constraints on the 3D magnetic field and density properties of Musca, which is directly linked to the resonant frequencies of linear MHD modes.

\subsection{Striations in Taurus}
\label{sec:taurus}

\begin{figure}
\begin{center}
\begin{tabular}{c}
\includegraphics[height=7.5cm]{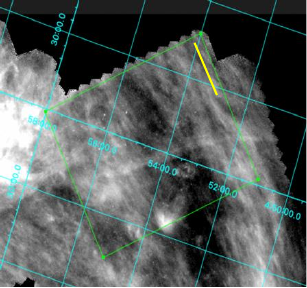}
\end{tabular}
\end{center}
\caption 
{\label{fig:taurus} Striations in Taurus. Striations likely channel material towards a denser core (not shown in the image). Colored shapes have the same meaning as in Fig.~\ref{fig:polaris_flare}.} 
\end{figure} 

The kinematic properties of striations in the Taurus molecular cloud, as derived from spectroscopic observation of CO lines, are consistent with the linear MHD theory. Several wave modes have been detected there in the power spectra, indicating that wave periods range from 0.5 - 1.5 Myrs \cite{tritsis_2018.new.method}. Employing an Alfv\'en speed close to 1 km/s, we obtain that the shortest wavelength of these modes should be $\sim 0.5$ pc, which can be probed using any band of PRIMA. PRIMAger's polarization imaging and its high resolution will enable the power spectrum analysis of the polarization angles in the target region (Fig.~\ref{fig:taurus}), allowing us to even search for high-frequency wave modes that were previously undetectable.

Fluctuations in the polarization angles are mostly induced by incompressible wave modes, while column density fluctuations are induced by compressible waves. By constraining the relative frequencies of the various modes, we can estimate the relative energy ratio between compressible and incompressible waves. Synergies between PRIMAger's polarization data and spectroscopic data from ground-based radio observatories could provide important constraints on the energy ratios of MHD waves, which is an important parameter of turbulent cloud simulations \cite{tritsis_2025.numerical.sims.proj}.

\begin{table}[h!]
\centering
\begin{tabular}{|l|c|c|c|c|c|c|c|c|c|}
\hline
 & & & \multicolumn{3}{c|}{\textbf{Intensities}} & \multicolumn{2}{c|}{\textbf{SED parameters}}\\
\hline
 \textbf{Cloud} & \textbf{R.A.} & \textbf{Dec.}  & \textbf{250 $\mu$m} & \textbf{350 $\mu$m} & \textbf{500 $\mu$m} & $\beta$ & $T_d$\\
\hline
Polaris Flare & 05:46:19 & +87:57:12 & 7.3 &  3.6 & 1.5 & 2.9 & 14 \\
Musca & 12:26:24 & -71:59:48 &   32.8 & 19.6 & 8.4 & 2.0 & 15 \\
Taurus & 04:54:31 & +27:00:44 &  36.0 & 22.5 & 9.9 & 2.0 & 15\\
\hline
\end{tabular}
\vspace{0.25cm}
\caption{Central coordinates and main properties of the target regions used for estimating the exposure times. The intensities are in units of MJy/sr and were measured using the SPIRE instrument on the Herschel satellite. The SED parameters were obtained from previous works (Polaris Flare\cite{Miville-Deschenes10}, Musca\cite{cox_2016},and Taurus\cite{flagey_2009.dust.temperature.taurus}).}
\label{table:herschel_intensities}
\end{table}

\begin{table}[h!]
\centering
\begin{tabular}{|l|c|c|c|c|c|c|c|c|}
\hline
& \multicolumn{4}{c|}{\textbf{Intensities} (MJy/sr)} & \multicolumn{4}{c|}{\textbf{Polarized Intensities} (MJy/sr)}\\
\hline
 \textbf{Cloud} & \textbf{92 $\mu$m} & \textbf{126 $\mu$m} &  \textbf{172 $\mu$m} & \textbf{235 $\mu$m} & \textbf{92 $\mu$m} & \textbf{126 $\mu$m} &  \textbf{172 $\mu$m} & \textbf{235 $\mu$m} \\
\hline
Polaris Flare & 2.2 & 6.7 & 9.8 &  7.8 &  0.033 & 0.101 &	0.147 & 0.117 \\
Musca & 5.0 & 19.5 & 33.9 & 33.97 & 0.20 & 0.78 & 1.36 & 1.36 \\
Taurus & 7.1 & 25.1 & 41.3 & 39.2 & 0.14 & 0.50 &	0.83 & 0.78 \\
\hline
\end{tabular}
\vspace{0.25cm}
\caption{Estimated total and polarized fluxes for the various bands of PRIMAger towards the target regions.}
\label{table:polarization_flux_estimates}
\end{table}

\begin{table}[h!]
\centering
\begin{tabular}{|l|c|c|c|c|c|c|c|c|}
\hline
 &  & \multicolumn{4}{c|}{\textbf{Estimated Exposure Time (hours)}} \\
\hline
 \textbf{Cloud} & \textbf{Sensitivity} & \textbf{92 $\mu$m} & \textbf{126 $\mu$m} &  \textbf{172 $\mu$m} & \textbf{235 $\mu$m}  \\
\hline
Polaris Flare & $5\sigma$ & 1942.4 & 200.0 & 28.7 & 24.5
\\
& $10\sigma$ & 7769.6 & 457.5 & 114.9 & 98.2
\\
\hline
Musca & $5 \sigma$ & 53.1  & 1.9 & 0.3 & 0.2 \\
      & $10 \sigma$& 212.5 & 7.6 & 1.4 & 0.7  \\
      & $20 \sigma$& 849.9 & 30.4	& 5.4 & 2.9  \\
\hline
Taurus & $5 \sigma$ & 220.6 & 4.6 & 0.9 & 0.5 \\
      & $10 \sigma$& 882.6  & 18.3 & 3.7 & 2.2 \\
      & $20 \sigma$& 3527.7 & 73.4 & 14.7 & 8.8 \\
      \hline
\end{tabular}
\vspace{0.25cm}
\caption{Estimated exposure times for PRIMAger's bands at various sensitivity depths.}
\label{table:exp_time}
\end{table}

\subsection{Exposure time estimation}
\label{sec:exp_time}

For the exposure time calculation we employed the following assumptions. We assumed that dust grains emit as gray body with a spectral energy distribution (SED) given by\cite{hildebrand_1983.cloud.masses}: $I_\nu = \epsilon_\nu (\beta) B_\nu (T_d)$, where $\epsilon_\nu \propto \nu^\beta$ is the frequency-dependent emissivity of dust, $B_\nu$ is the black body function, and $T_d$ is the dust temperature. We used existing constraints for $\beta$ and $T_d$ in the Polaris Flare\cite{Miville-Deschenes10}, Taurus\cite{flagey_2009.dust.temperature.taurus}, and Musca\cite{cox_2016}. Table~\ref{table:herschel_intensities} shows the central coordinates of the proposed regions along with the employed SED parameters. For reference, we also report the average intensities at $250 \mu$m, $350 \mu$m, and $500 \mu$m of each target region, as measured by the SPIRE instrument onboard the Herschel satellite.

We estimated the total intensity flux of each region at the four bands of PRIMAger using the analytical expression for the SED. To estimate the polarized fluxes in each region, we utilized starlight polarization. In the striations region of the Polaris Flare, the degree of polarization ($p$), as traced by starlight data, is $\sim 1.5 \%$ \cite{panopoulou_2016.PFstriations}. The degree of polarization of dust emission is independent of frequency in the wavelength range $100 - 1000 ~\mu$m \cite{Hensley23}. Thus, we can estimate the polarized flux at any PRIMA frequency ($\nu$) assuming a constant $p$. We obtain the following relation for the polarization fluxes, $P_\nu = p I_\nu$.  We estimated the polarization fluxes in every cloud using this relation and archival starlight polarization data.

For Musca, we obtain from starlight data that $p \sim 4\%$ \cite{pereyra_2004.musca}. We note that there are not any starlight polarization data towards the dense part of Musca. The degree of polarization is expected to decrease there (Fig.~\ref{fig:musca}) either due to enhanced tangling of the field lines \cite{kandori_2022} or suppression of grain alignment efficiency \cite{lazarian_2007.rat.theory}. The expected decrease in the degree of polarization in the dense part of Musca would be compensated by the enhanced column density. For this reason, we expect the polarized intensities in the striations and the dense part of Musca to be roughly similar. In Taurus, starlight polarization measurements have been made at near-infrared wavelengths ($\sim 1 \mu$m), where the degree of polarization is lower than optical \cite{Serkowski73}. Thus, we treat the observed average value, which is $\sim 2 \%$ \cite{Chapman11}, as a lower limit. The estimated average total intensity and polarized intensity for the various bands of PRIMAger are shown in Table~\ref{table:polarization_flux_estimates}.

Table~\ref{table:exp_time} shows the estimated exposure times of the various PRIMAger bands at different sensitivities. In the Polaris Flare, we wish to search for the sausage mode oscillation pattern in the polarization angle map. The target accuracy in the polarization angle is $\sim 1^o$ (\S~\ref{sec:polaris_flare}), which requires SNR $\sim 20$. Because such a deep survey would require long integration times, we propose the following strategy to detect the expected signal within a reasonable time. The wavelength on the sky of the candidate sausage mode pattern in the Polaris Flare is $\sim 1.3^o$ (Fig.~\ref{fig:polaris_flare}), which corresponds to $\sim 9$ pc. This means that at the 27.6$''$ resolution of PRIMAger, we would obtain $\sim 170$ independent measurements. Decreasing the resolution by a factor of 4, through spatial averaging, would increase the SNR by a factor of 2, while the number of independent measurements (in this case 43) would still be sufficient to probe the magnetic field fluctuations. Thus, a 10$\sigma$ detection in averaged polarized emission would be sufficient to probe the magnetic field properties in the striations of the Polaris Flare. This would require $\sim 98$ hours with band 4 of PRIMAger (Table~\ref{table:exp_time}).

To further reduce the integration times, we can combine measurements from multiple bands. If, for simplicity, we assume that the SNR is similar in bands 3 and 4 at a given exposure, then averaging the polarization angle measurements from the two bands would increase the SNR by a factor of $\sqrt{2}$. The combination of the two averaging methods implies that we would need a $7 \sigma$-deep survey to reach a SNR of $\sim 20$. This can be achieved within 57.5, and 49.1 hours for bands 3 and 4, respectively. Since all bands are measured simultaneously, we propose for $\sim 57.5$ hours (Table~\ref{table:exp_time}).

To motivate the averaging between the two PRIMA bands, we inspected the various SPIRE images, which showed a tight correlation of the dust structures in the various bands. As a result, the polarization measurements at the various PRIMA bands would trace the same magnetic field morphology, which means that no intrinsic differences are expected in the polarization angles between bands 2 and 3 of PRIMA in this cloud.

Dust grains in the striations of Musca and Taurus are generally expected to exhibit a well-defined mean temperature, except in the dense part of Musca, where both small hot and large cold grains likely contribute to the continuum emission. In this case, measurements may show frequency dependence, due to dust temperature variations. We estimate that achieving a $5 \sigma$ accuracy in the average polarization flux of Musca in bands 2, 3, and 4 of PRIMAger would require $\sim 2$ hours. In Taurus, we estimate $5 \sigma$ sensitivity in bands 3 and 4 within an hour. Overall, we estimate a total of $\sim 59$ hours to observe all three regions with PRIMAger, with most of the exposure time allocated to the Polaris Flare, which displays the most promising sausage mode oscillatory pattern (Fig.~\ref{fig:polaris_flare}).

\section{Discussion \& Conclusions}

In this article, we have outlined a comprehensive observational strategy to investigate the properties of striations in three nearby molecular clouds -- the Polaris Flare, Taurus, and Musca -- using the PRIMA polarimeter. All these regions show strong evidence of MHD waves propagation. Our primary goal is to observationally test specific predictions of the linear MHD theory.

The proposed survey program includes three well-studied molecular clouds with prominent striations, but this choice of clouds is by no means restrictive. Striations may be found throughout our Galaxy, as the ISM is a highly dynamic environment -- constantly perturbed, hence continuously exciting MHD waves. To preserve their quasi-periodicity and linearity, which distinguish them from other ISM structures, striations need to be relatively isolated and unaffected by star formation activity. For example, the L914 dark cloud exhibits well-defined striations, which satisfy the dispersion relation of MHD waves \cite{sun_2024.striations.L914.cloud}. PRIMAger data would be valuable for the characterization of the properties of striations in this region. 

PRIMAger is uniquely suited for the proposed program because previous observatories (e.g., Planck and SOFIA) lacked the required resolution and sensitivity. Current observatories, such as the James Clerk Maxwell Telescope, and upcoming ones, like the Fred Young Submillimeter Telescope (FYST), are less sensitive to the diffuse dust emission which typically occurs at dust temperatures (15 - 20 K) and necessitates short-wavelength ($\sim 100 \mu$m) observations. FYST will feature a $\sim 200 \mu$m band capable of tracing diffuse dust structures. However, two of our main targets -- the Polaris Flare and Taurus -- are barely visible from its site.

Given their quiescent nature, striations are ideal testbeds to study linear MHD waves. The detection of specific oscillatory patterns in the ISM, such as sausage modes, would provide significant conceptual links between Solar and ISM turbulence, being a major leap for the latter which lacks of concrete observational constraints. Such a detection would be direct evidence of the existence of coherent magnetic flux structures in the ISM, which is a fundamental element of compressible MHD theories \cite{skalidis_2023.theory.coherent.structures}. PRIMAger is the only observatory suited for the hunt of linear MHD mode oscillations in the ISM. The proposed observations require a total of $\sim 59$ hours and promise to fill critical gaps in our theoretical understanding of the early stages of star formation and the role of magnetic fields in shaping the ISM. 

\subsection*{Disclosures}

The authors declare there are no financial interests, commercial affiliations, or other potential conflicts of interest that have influenced the objectivity of this research or the writing of this paper.

\subsection*{Code, Data, and Materials Availability}

The preparation of this manuscript included archival data that are publicly available, as well as numerical data from simulations conducted by the authors. To access the numerical simulations, please contact the corresponding author.

\subsection*{Acknowledgments}

We thank the two anonymous reviewers for their constructive feedback, which significantly improved the manuscript. We are grateful to P. Kallemi for her valuable assistance with figures. Support for this work was provided by NASA through the NASA Hubble Fellowship grant HST-HF2-51566.001 awarded by the Space Telescope Science Institute, which is operated by the Association of Universities for Research in Astronomy, Inc., for NASA, under contract NAS5-26555. A. Tritsis acknowledges support by the Ambizione grant no. PZ00P2\_202199 of the Swiss National Science Foundation (SNSF). This work was performed in part at the Jet Propulsion Laboratory, California Institute of Technology, under contract with the National Aeronautics and Space Administration (80NM0018D0004).

\bibliography{report}   

\begin{thebibliography}{10}

\bibitem{valdivia_2016.H2.formation.simulations}
V.~{Valdivia}, P.~{Hennebelle}, M.~{G{\'e}rin}, {\em et~al.}, ``{H$_{2}$
  distribution during the formation of multiphase molecular clouds},'' {\em
  Astronomy \& Astrophysics} {\bf 587}, A76  (2016).

\bibitem{clarck_2014.hi.fibers}
S.~E. {Clark}, J.~E.~G. {Peek}, and M.~E. {Putman}, ``{Magnetically Aligned H I
  Fibers and the Rolling Hough Transform},'' {\em ApJ} {\bf 789}, 82  (2014).

\bibitem{kalberla_2016.hi.fibers}
P.~M.~W. {Kalberla}, J.~{Kerp}, U.~{Haud}, {\em et~al.}, ``{Cold Milky Way HI
  Gas in Filaments},'' {\em ApJ} {\bf 821}, 117  (2016).

\bibitem{Bracco20}
A.~Bracco, V.~Jeli{\'c}, A.~Marchal, {\em et~al.}, ``The multiphase and
  magnetized neutral hydrogen seen by {LOFAR},'' {\em Astronomy \&
  Astrophysics} {\bf 644}, L3  (2020).

\bibitem{skalidis_2022.ursa.major}
R.~{Skalidis}, K.~{Tassis}, G.~V. {Panopoulou}, {\em et~al.},
  ``{H$_{I}$-H$_{2}$ transition: Exploring the role of the magnetic field. A
  case study toward the Ursa Major cirrus},'' {\em Astronomy \& Astrophysics}
  {\bf 665}, A77  (2022).

\bibitem{planck_xxxv_2016}
{Planck Collaboration}, P.~A.~R. {Ade}, N.~{Aghanim}, {\em et~al.}, ``{Planck
  intermediate results. XXXV. Probing the role of the magnetic field in the
  formation of structure in molecular clouds},'' {\em Astronomy \&
  Astrophysics} {\bf 586}, A138  (2016).

\bibitem{skalidis_2019}
R.~{Skalidis} and V.~{Pelgrims}, ``{Local Bubble contribution to the 353-GHz
  dust polarized emission},'' {\em AStronomy \& Astrophysics} {\bf 631}, L11
  (2019).

\bibitem{glenn_2024.prima.concept}
J.~Glenn, M.~Meixner, C.~M. Bradford, {\em et~al.}, ``{PRIMA: the probe
  far-infrared mission for astrophysics},'' in {\em Space Telescopes and
  Instrumentation 2024: Optical, Infrared, and Millimeter Wave},  L.~E. Coyle,
  S.~Matsuura, and M.~D. Perrin, Eds.,  {\bf 13092}, 130920J, International
  Society for Optics and Photonics, SPIE  (2024).

\bibitem{Andre10}
P.~Andr{\'e}, A.~Men'shchikov, S.~Bontemps, {\em et~al.}, ``From filamentary
  clouds to prestellar cores to the stellar {IMF}: {Initial} highlights from
  the {Herschel} {Gould} {Belt} {Survey},'' {\em Astronomy \& Astrophysics}
  {\bf 518}, L102  (2010).

\bibitem{panopoulou_2017.filament.width}
G.~V. {Panopoulou}, I.~{Psaradaki}, R.~{Skalidis}, {\em et~al.}, ``{A closer
  look at the `characteristic' width of molecular cloud filaments},'' {\em
  MNRAS} {\bf 466}, 2529--2541  (2017).

\bibitem{hacar_2023.filament.review}
A.~{Hacar}, S.~E. {Clark}, F.~{Heitsch}, {\em et~al.}, ``{Initial Conditions
  for Star Formation: a Physical Description of the Filamentary ISM},'' in {\em
  Protostars and Planets VII},  S.~{Inutsuka}, Y.~{Aikawa}, T.~{Muto}, {\em
  et~al.}, Eds., {\em Astronomical Society of the Pacific Conference Series}
  {\bf 534}, 153  (2023).

\bibitem{konyves_2015}
V.~{K{\"o}nyves}, P.~{Andr{\'e}}, A.~{Men'shchikov}, {\em et~al.}, ``{A census
  of dense cores in the Aquila cloud complex: SPIRE/PACS observations from the
  Herschel Gould Belt survey},'' {\em Astronomy \& Astrophysics} {\bf 584}, A91
   (2015).

\bibitem{federrath_2021}
C.~{Federrath}, R.~S. {Klessen}, L.~{Iapichino}, {\em et~al.}, ``{The sonic
  scale of interstellar turbulence},'' {\em Nature Astronomy} {\bf 5}, 365--371
   (2021).

\bibitem{inoue_2018.filament.formation}
T.~{Inoue}, P.~{Hennebelle}, Y.~{Fukui}, {\em et~al.}, ``{The formation of
  massive molecular filaments and massive stars triggered by a
  magnetohydrodynamic shock wave},'' {\em PASJ} {\bf 70}, S53  (2018).

\bibitem{Hennebelle13}
P.~{Hennebelle} and P.~{Andr{\'e}}, ``{Ion-neutral friction and
  accretion-driven turbulence in self-gravitating filaments},'' {\em Astronomy
  \& Astrophysics} {\bf 560}, A68  (2013).

\bibitem{Goldsmith08}
P.~F. {Goldsmith}, M.~{Heyer}, G.~{Narayanan}, {\em et~al.}, ``{Large-Scale
  Structure of the Molecular Gas in Taurus Revealed by High Linear Dynamic
  Range Spectral Line Mapping},'' {\em The Astrophysical Journal} {\bf 680},
  428--445  (2008).

\bibitem{tritsis_2019}
A.~{Tritsis}, C.~{Federrath}, and V.~{Pavlidou}, ``{Magnetic Field Tomography
  in Two Clouds toward Ursa Major Using H I Fibers},'' {\em ApJ} {\bf 873}, 38
  (2019).

\bibitem{Boulanger88}
F.~{Boulanger} and M.~{Perault}, ``{Diffuse Infrared Emission from the Galaxy.
  I. Solar Neighborhood},'' {\em The Astrophysical Journal} {\bf 330}, 964
  (1988).

\bibitem{Miville-Deschenes10}
M.~A. {Miville-Desch{\^e}nes}, P.~G. {Martin}, A.~{Abergel}, {\em et~al.},
  ``{Herschel-SPIRE observations of the Polaris flare: Structure of the diffuse
  interstellar medium at the sub-parsec scale},'' {\em Astronomy \&
  Astrophysics} {\bf 518}, L104  (2010).

\bibitem{Hennemann12}
M.~{Hennemann}, F.~{Motte}, N.~{Schneider}, {\em et~al.}, ``{The spine of the
  swan: a Herschel study of the DR21 ridge and filaments in Cygnus X},'' {\em
  Astronomy \& Astrophysics} {\bf 543}, L3  (2012).

\bibitem{Palmeirim13}
P.~{Palmeirim}, P.~{Andr{\'e}}, J.~{Kirk}, {\em et~al.}, ``{Herschel view of
  the Taurus B211/3 filament and striations: evidence of filamentary
  growth?},'' {\em Astronomy \& Astrophysics} {\bf 550}, A38  (2013).

\bibitem{AlvesdeOliveira14}
C.~{Alves de Oliveira}, N.~{Schneider}, B.~{Mer{\'\i}n}, {\em et~al.},
  ``{Herschel view of the large-scale structure in the Chamaeleon dark
  clouds},'' {\em Astronomy \& Astrophysics} {\bf 568}, A98  (2014).

\bibitem{Chapman11}
N.~L. {Chapman}, P.~F. {Goldsmith}, J.~L. {Pineda}, {\em et~al.}, ``{The
  Magnetic Field in Taurus Probed by Infrared Polarization},'' {\em The
  Astrophysical Journal} {\bf 741}, 21  (2011).

\bibitem{panopoulou_2016.PFstriations}
G.~V. {Panopoulou}, I.~{Psaradaki}, and K.~{Tassis}, ``{The magnetic field and
  dust filaments in the Polaris Flare},'' {\em MNRAS} {\bf 462}, 1517--1529
  (2016).

\bibitem{Heyer16}
M.~{Heyer}, P.~F. {Goldsmith}, U.~A. {Y{\i}ld{\i}z}, {\em et~al.},
  ``{Striations in the Taurus molecular cloud: Kelvin-Helmholtz instability or
  MHD waves?},'' {\em MNRAS} {\bf 461}, 3918--3926  (2016).

\bibitem{Tritsis16}
A.~{Tritsis} and K.~{Tassis}, ``{Striations in molecular clouds: streamers or
  MHD waves?},'' {\em MNRAS} {\bf 462}, 3602--3615  (2016).

\bibitem{Chen17}
C.-Y. {Chen}, Z.-Y. {Li}, P.~K. {King}, {\em et~al.}, ``{Fantastic Striations
  and Where to Find Them: The Origin of Magnetically Aligned Striations in
  Interstellar Clouds},'' {\em The Astrophysical Journal} {\bf 847}, 140
  (2017).

\bibitem{skalidis_2023.striations}
R.~{Skalidis}, K.~{Gkimisi}, K.~{Tassis}, {\em et~al.}, ``{CO enhancement by
  magnetohydrodynamic waves. Striations in the Polaris Flare},'' {\em Astronomy
  \& Astrophysics} {\bf 673}, A76  (2023).

\bibitem{Li13}
H.-b. {Li}, M.~{Fang}, T.~{Henning}, {\em et~al.}, ``{The link between magnetic
  fields and filamentary clouds: bimodal cloud orientations in the Gould
  Belt},'' {\em MNRAS} {\bf 436}, 3707--3719  (2013).

\bibitem{Nakariakov16}
V.~M. {Nakariakov}, V.~{Pilipenko}, B.~{Heilig}, {\em et~al.},
  ``{Magnetohydrodynamic Oscillations in the Solar Corona and Earth's
  Magnetosphere: Towards Consolidated Understanding},'' {\em Space Science
  Reviews} {\bf 200}, 75--203  (2016).

\bibitem{Tritsis18}
A.~{Tritsis} and K.~{Tassis}, ``{Magnetic seismology of interstellar gas
  clouds: Unveiling a hidden dimension},'' {\em Science} {\bf 360}, 635--638
  (2018).

\bibitem{ciesla_2024.primager}
L.~Ciesla, D.~Burgarella, C.~D. Dowell, {\em et~al.}, ``{PRIMA: PRIMAger, a far
  infrared hyperspectral and polarimetric instrument},'' in {\em Space
  Telescopes and Instrumentation 2024: Optical, Infrared, and Millimeter Wave},
   L.~E. Coyle, S.~Matsuura, and M.~D. Perrin, Eds.,  {\bf 13092}, 130920K,
  International Society for Optics and Photonics, SPIE  (2024).

\bibitem{burgarella_2024.primager}
D.~Burgarella, L.~Ciesla, M.~Sauvage, {\em et~al.}, ``{PRIMA: science cases and
  requirements for the photometric and polarimetric PRIMAger far-infrared
  camera},'' in {\em Space Telescopes and Instrumentation 2024: Optical,
  Infrared, and Millimeter Wave},  L.~E. Coyle, S.~Matsuura, and M.~D. Perrin,
  Eds.,  {\bf 13092}, 130923B, International Society for Optics and Photonics,
  SPIE  (2024).

\bibitem{yang_2019.energy.occupation}
L.~P. {Yang}, H.~{Li}, S.~T. {Li}, {\em et~al.}, ``{Energy occupation of waves
  and structures in 3D compressive MHD turbulence},'' {\em MNRAS} {\bf 488},
  859--867  (2019).

\bibitem{goldreich_sridhar_1995}
P.~{Goldreich} and S.~{Sridhar}, ``{Toward a Theory of Interstellar Turbulence.
  II. Strong Alfvenic Turbulence},'' {\em ApJ} {\bf 438}, 763  (1995).

\bibitem{boldyrev_2006.dynamic.alignment}
S.~{Boldyrev}, ``{Spectrum of Magnetohydrodynamic Turbulence},'' {\em PhRvL}
  {\bf 96}, 115002  (2006).

\bibitem{roberts_1984.sausage.modes}
B.~{Roberts}, P.~M. {Edwin}, and A.~O. {Benz}, ``{On coronal oscillations},''
  {\em ApJ} {\bf 279}, 857--865  (1984).

\bibitem{nakariakov_2003.sausage.first.detection}
V.~M. {Nakariakov}, V.~F. {Melnikov}, and V.~E. {Reznikova}, ``{Global sausage
  modes of coronal loops},'' {\em Astronomy \& Astrophysics} {\bf 412}, L7--L10
   (2003).

\bibitem{king_2014.robopol}
O.~G. {King}, D.~{Blinov}, A.~N. {Ramaprakash}, {\em et~al.}, ``{The RoboPol
  pipeline and control system},'' {\em MNRAS} {\bf 442}, 1706--1717  (2014).

\bibitem{sousbie_2011.disperse.algorithm}
T.~{Sousbie}, ``{The persistent cosmic web and its filamentary structure - I.
  Theory and implementation},'' {\em MNRAS} {\bf 414}, 350--383  (2011).

\bibitem{panopoulou_2014.taurus.filaments}
G.~V. {Panopoulou}, K.~{Tassis}, P.~F. {Goldsmith}, {\em et~al.}, ``{$^{13}$CO
  filaments in the Taurus molecular cloud},'' {\em MNRAS} {\bf 444}, 2507--2524
   (2014).

\bibitem{King18}
P.~K. King, L.~M. Fissel, C.-Y. Chen, {\em et~al.}, ``Mnras, 474, 5122,'' {\em
  MNRAS} {\bf 474}, 5122  (2018).

\bibitem{WardThomson_2010.polaris.cores}
D.~{Ward-Thompson}, J.~M. {Kirk}, P.~{Andr{\'e}}, {\em et~al.}, ``{A Herschel
  study of the properties of starless cores in the Polaris Flare dark cloud
  region using PACS and SPIRE},'' {\em Astronomy \& Astrophysics} {\bf 518},
  L92  (2010).

\bibitem{myers_2017.musca}
P.~C. {Myers}, ``{Star-forming Filament Models},'' {\em ApJ} {\bf 838}, 10
  (2017).

\bibitem{hartmann_2002.taurus.sfr}
L.~{Hartmann}, ``{Flows, Fragmentation, and Star Formation. I. Low-Mass Stars
  in Taurus},'' {\em ApJ} {\bf 578}, 914--924  (2002).

\bibitem{bonne_2020.musca.co.obs}
L.~{Bonne}, S.~{Bontemps}, N.~{Schneider}, {\em et~al.}, ``{Formation of the
  Musca filament: evidence for asymmetries in the accretion flow due to a
  cloud-cloud collision},'' {\em Astronomy \& AStrophysics} {\bf 644}, A27
  (2020).

\bibitem{panopoulou_2022.cloud.distances}
G.~V. {Panopoulou}, S.~E. {Clark}, A.~{Hacar}, {\em et~al.}, ``{The width of
  Herschel filaments varies with distance},'' {\em Astronomy \& Astrophysics}
  {\bf 657}, L13  (2022).

\bibitem{grant_2015}
S.~D.~T. {Grant}, D.~B. {Jess}, M.~G. {Moreels}, {\em et~al.}, ``{Wave Damping
  Observed in Upwardly Propagating Sausage-mode Oscillations Contained within a
  Magnetic Pore},'' {\em ApJ} {\bf 806}, 132  (2015).

\bibitem{edwin_roberts1983.wave.propagation.magnetic.cylinders}
P.~M. {Edwin} and B.~{Roberts}, ``{Wave Propagation in a Magnetic Cylinder},''
  {\em Sol. Phys.} {\bf 88}, 179--191  (1983).

\bibitem{tritsis_2018.new.method}
A.~{Tritsis}, C.~{Federrath}, N.~{Schneider}, {\em et~al.}, ``{A new method for
  probing magnetic field strengths from striations in the interstellar
  medium},'' {\em MNRAS} {\bf 481}, 5275--5285  (2018).

\bibitem{Skalidis2021High-accuracyPolarization}
R.~{Skalidis} and K.~{Tassis}, ``{High-accuracy estimation of magnetic field
  strength in the interstellar medium from dust polarization},'' {\em Astronomy
  \& Astrophysics} {\bf 647}, A186  (2021).

\bibitem{skalidis_2021.sqrt}
R.~{Skalidis}, J.~{Sternberg}, J.~R. {Beattie}, {\em et~al.}, ``{Why take the
  square root? An assessment of interstellar magnetic field strength estimation
  methods},'' {\em Astronomy \& Astrophysics} {\bf 656}, A118  (2021).

\bibitem{cox_2016}
N.~L.~J. {Cox}, D.~{Arzoumanian}, P.~{Andr{\'e}}, {\em et~al.}, ``{Filamentary
  structure and magnetic field orientation in Musca},'' {\em Astronomy \&
  Astrophysics} {\bf 590}, A110  (2016).

\bibitem{tritsis_2022.musca}
A.~{Tritsis}, F.~{Bouzelou}, R.~{Skalidis}, {\em et~al.}, ``{The musca
  molecular cloud: The perfect 'filament' is still a sheet},'' {\em MNRAS} {\bf
  514}, 3593--3603  (2022).

\bibitem{andre_2019}
P.~{Andr{\'e}}, A.~{Hughes}, V.~{Guillet}, {\em et~al.}, ``{Probing the cold
  magnetised Universe with SPICA-POL (B-BOP)},'' {\em PASA} {\bf 36}, e029
  (2019).

\bibitem{tritsis_2025.numerical.sims.proj}
A.~{Tritsis}, S.~{Basu}, and C.~{Federrath}, ``{Projection-angle effects when
  ``observing'' a turbulent magnetized collapsing molecular cloud: I. Chemistry
  and line transfer},'' {\em AStronomy \& Astrophysics} {\bf 695}, A18  (2025).

\bibitem{flagey_2009.dust.temperature.taurus}
N.~{Flagey}, A.~{Noriega-Crespo}, F.~{Boulanger}, {\em et~al.}, ``{Evidence for
  Dust Evolution Within the Taurus Complex from Spitzer Images},'' {\em ApJ}
  {\bf 701}, 1450--1463  (2009).

\bibitem{hildebrand_1983.cloud.masses}
R.~H. {Hildebrand}, ``{The determination of cloud masses and dust
  characteristics from submillimetre thermal emission.},'' {\em QJRAS} {\bf
  24}, 267--282  (1983).

\bibitem{Hensley23}
B.~S. Hensley and B.~T. Draine, ``The astrodust+pah model: A unified
  description of the extinction, emission, and polarization from dust in the
  diffuse interstellar medium,'' {\em The Astrophysical Journal} {\bf 948}, 55
  (2023).

\bibitem{pereyra_2004.musca}
A.~{Pereyra} and A.~M. {Magalh{\~a}es}, ``{Polarimetry toward the Musca Dark
  Cloud. I. The Catalog},'' {\em ApJ} {\bf 603}, 584--594  (2004).

\bibitem{kandori_2022}
R.~{Kandori}, M.~{Tamura}, M.~{Saito}, {\em et~al.}, ``{Distortion of magnetic
  fields in Barnard 68},'' {\em PASJ} {\bf 72}, 8  (2020).

\bibitem{lazarian_2007.rat.theory}
A.~{Lazarian} and T.~{Hoang}, ``{Radiative torques: analytical model and basic
  properties},'' {\em MNRAS} {\bf 378}, 910--946  (2007).

\bibitem{Serkowski73}
K.~Serkowski, ``In interstellar dust and related topics, ed. jm greenberg, hc
  van de hulst. iau symp. 52:145,'' in {\em IAU Symp. 52:145},  (Dordrect,
  Neth.: Kluwer)  (1973).

\bibitem{sun_2024.striations.L914.cloud}
L.~{Sun}, X.~{Chen}, M.~{Fang}, {\em et~al.}, ``{Magnetically Aligned
  Striations in the L914 Filamentary Cloud},'' {\em ApJ} {\bf 167}, 176
  (2024).

\bibitem{skalidis_2023.theory.coherent.structures}
R.~{Skalidis}, K.~{Tassis}, and V.~{Pavlidou}, ``{Analytic characterization of
  sub-Alfv{\'e}nic turbulence energetics},'' {\em Astronomy \& Astrophysics}
  {\bf 672}, L3  (2023).

\end{thebibliography}
\bibliographystyle{spiejour}   

\vspace{2ex}\noindent\textbf{Raphael Skalidis} is a NASA Hubble Postdoctoral Fellow at the California Institute of Technology. He received his PhD in physics from the University of Crete in 2022. His research is focused on interstellar medium magnetic fields. 

\vspace{2ex}\noindent\textbf{Konstantinos Tassis} is a professor of theoretical astrophysics at the Department of Physics of the University of Crete in Greece, and affiliated faculty of the Institute of Astrophysics of the Foundation for Research and Technology - Hellas. He is the management panel chair of the PASIPHAE Collaboration. His research focus is interstellar medium physics and star formation, with an emphasis on the role of magnetic fields.  

\vspace{2ex}\noindent\textbf{Aris Tritsis} is an Ambizione Postdoctoral Fellow at Ecole Polytechnique F\'ed\'erale de Lausanne. He received his PhD in physics from the University of Crete in 2017. He studies the evolution and observational properties of interstellar molecular clouds through 3-dimensional (non-)ideal MHD and radiative transfer simulations.

\vspace{2ex}\noindent\textbf{Paul F. Goldsmith} is a Senior Research Scientist at JPL.  He is involved in the NASA balloon missions GUSTO and ASTHROS to observe fine structure line emission probing feedback from massive star formation.  He studies formation and evolution of interstellar molecular clouds through millimeter, submillimeter and infrared observations.

\listoffigures
\listoftables

\end{spacing}
\end{document}